# Low-temperature specific heat of graphite and CeSb$_2$: Validation of a quasi-adiabatic continuous method

T. Pérez-Castañeda, J. Azpeitia, J. Hanko, A. Fente, H. Suderow, and M. A. Ramos

*Laboratorio de Bajas Temperaturas, Departamento de Física de la Materia Condensada and Instituto de Ciencia de Materiales "Nicolás Cabrera" Universidad Autónoma de Madrid, Cantoblanco, E-28049 Madrid SPAIN*

Tel.: +34 91 497 4756

Fax: +34 91 497 3961

E-mail: tomas.perez@uam.es

We present the application of a fast *quasi-adiabatic* continuous method to the measurement of specific heat at $^4$He temperatures, which can be used for the study of a wide range of materials. The technique can be performed in the same configuration used for the relaxation method, as the typical time constants between calorimetric cell and thermal sink at 4.2 K are chosen to be of the order of $\tau \sim 30$ s. The accuracy in the absolute values have been tested by comparing them to relaxation-method results obtained in the same samples (performed *in situ* using the same set-up), with a deviation between the absolute values < 3% in the whole temperature range. This new version of the continuous calorimetric method at low temperatures allows us to completely characterize and measure a sample within a few hours with a high density of data points, whereas when employing other methods we typically need a few days. An exhaustive study has been performed for reproducibility to be tested. In the present work, we have applied this method to two different substances: CeSb$_2$, which exhibits three magnetic transitions at 15.5 K, 11.7 K and 9.5 K, and graphite, both highly-oriented pyrolytic graphite (HOPG) and natural crystals. Our results on these graphites are discussed in comparison with previous published data on different kinds of graphite samples.

# 1 Introduction

Specific-heat measurements are a powerful tool for the study of the physical properties and the characterization of phase transitions in a wide variety of materials, ranging from e.g. superconductors or magnetic solids to glasses [1-3]. Low-temperature calorimeters have always been a challenging task for experimentalists, as a huge reduction in thermal energy of a factor $>10^4$ occurs between room temperature and a few Kelvin. Radiation, heat exchange with the environment, the lack of thermal equilibrium or the need to efficiently cool down the cell space, conform some of the hardest problems to solve when working at $^4$He temperatures using calorimetric techniques [4–9]. The first historical method to measure the specific heat is the adiabatic one, where the measured increase in the temperature is directly produced by a known amount of heat supplied to the sample. To fulfill the adiabatic condition, an extremely good thermal isolation of the sample from its surroundings is required [5], what makes it in many cases unaffordable. To avoid this problem many other techniques have been developed, such as AC calorimetry [6], or the thermal relaxation method (either the standard one [5,7] or alternative versions as such described in [8]), which still provide accurate absolute values of the specific heat. Nevertheless, all these methods can be much time consuming, what sometimes makes it complicated to apply on a big set of samples.

A faster alternative is the so-called "continuous heating method" [5], where a steady-state heating is continuously added to the sample at constant rate or power $P$, and only the resulting temperature increase $T(t)$ is recorded. Ideally, this technique requires an immediate distribution of heat within the calorimetric cell and in the sample itself. Hence the heating rate should be chosen small enough to guarantee a uniform distribution of the temperature.

In the traditional continuous heating method, the ideal adiabatic limit is assumed and the system therefore obeys the simple equation $P = C_p \cdot (dT/dt)$. An opposite approach to the adiabatic continuous method is found in Ref. [9], where the sample is strongly coupled to the surroundings. This is realized through an array of Peltier elements of ideally infinite thermal conductance, which are able to measure the heat flow, and then the heat capacity of the device.

We have developed a new (*quasi-adiabatic*) version [8] of the continuous method, using a different approach to those two mentioned above, though likely closer to the former one. However, the non-perfect adiabaticity of the real calorimeter is physically taken into account by using a more realistic equation of heat. Furthermore, our versatile calorimetric set-up [8] allows us to choose between this quasi-adiabatic continuous method and two alternative thermal relaxation methods for low-temperature specific-heat measurements, using the very same experimental set-up.

In this work, we have used an adapted version of our quasi-adiabatic continuous method, which has been employed for the first time to the measurement of low-temperature specific heat, aiming to check and validate its possible application at liquid-helium temperatures. As we will show below, this calorimetric technique essentially provides a fast and continuous set of data points in a wide temperature range (typically 3 hours for 2 K ≤ T ≤ 40 K) , though maintaining a good accuracy and precision in the absolute values: < 3% deviation compared to the relaxation technique. The aim of this work is to discuss the set-up requirements to apply this method and to show to what extent this method is reliable for low-temperature specific heat measurements in different scenarios: (i) to study substances presenting phase transitions, where accurate determination of the transition temperature and high resolution of the curve around the singularities are needed; and (ii) to analyze the specific heat in substances with a smooth behavior in $C_p(T)$, but for which the precise temperature dependence of the specific heat is crucial. For this purpose, we have chosen two different substances to be measured, one of each type, and which indeed present interesting phenomena in the field of low-temperature condensed matter physics. These two selected substances are $CeSb_2$, which exhibits magnetic transitions at low temperatures, and graphite, in both HOPG and natural forms.

The family of light rare-earth diantimonides $RSb_2$ (R = La–Nd) shows an extraordinary rich variety of physical properties going from superconductivity in $LaSb_2$ to anisotropic ferromagnetism in $CeSb_2$ [10-12]. In the particular case of $CeSb_2$, three low-temperature anomalies at 15.5 K, 11.7 K and 9.3 K have been observed, which correspond to magnetic transitions at zero magnetic field. Transport and magnetization experiments in this compound manifest fingerprints of up to four different magnetic phases, some of which are metastable [11].

On the other hand, the interest in carbon-based materials has been renewed in the last twenty years with the discovery of fullerenes [12-13] and nanotubes [15], and more recently, the first clear observation of a single atomic layer of graphite (graphene) further located carbon-based compounds in the center of interest in science and technology [16]. Nevertheless, a lot of properties lack a deep understanding even in the physics of graphite, as the problem of clearly identifying the expected $T$ and $T^3$ dependences in the low-temperature specific heat above 1 K [17,18]. A thorough study on graphite samples with different degrees of crystalline order is deemed essential to clarify some of the yet unsolved aspects.

## 2 Experimental

### 2.1 Sample preparation

Two different types of graphite were measured. First, a highly-oriented pyrolytic graphite (HOPG) sample of 12mm × 12mm × 1mm, and mass = 252 mg, from Advanced Ceramics with a rocking curve width of 0.4° (grade A). Using particle-induced x-ray emission (PIXE), the impurity content of these graphite samples were measured [19] and found to be negligible, with a concentration of magnetic impurities of the order of one magnetic atom per $10^6$ carbon atoms. For comparison, we have also measured two pieces of a sample of natural graphite (total mass = 292 mg), expected to have an even higher crystalline quality but also with a higher amount of impurity, however unknown.

The $CeSb_2$ sample has been grown in our laboratory from excess flux of antimony [20] using a stoichiometric mixture of 10% cerium and 90% antimony. The mixture was sealed and heated up to 950°C in 3 hours, kept at 950°C for another 3 hours and slowly cooled to 670°C in 70 hours. The growth produced large single crystals with plates of several millimeters.

### 2.2 Set-up

The experimental set-up is made up of a 25 mm in diameter and 0.5 mm thick sapphire disc, with mass m ≈ 1 g, onto which the sample, the thermometer and the heater are attached. A cooper wire 0.07 mm in diameter and 25 cm in length (used as the thermal link between calorimeter and reservoir) is glued to the sapphire forming a triangle with heater and thermometer. The sapphire disc is placed in the middle of a copper ring (reservoir) hanging with nylon wires. Typical internal equilibrium times of all the elements in the addenda-sample system are a few tens of milliseconds at liquid helium temperatures. Relaxation-time constants between the calorimetric cell and the thermal sink (copper ring) at those temperatures is $\tau \approx 30$ s, this is $10^3$–$10^4$ slower than the internal equilibration constants. A double-chamber insert is used to independently control the reservoir temperature (see Fig 1). The inner chamber is under high vacuum conditions $P \leq 10^{-7}$ mbar, whereas the outer chamber is under low helium gas pressure ($P \approx 1$ mbar at room temperature).

Two experimental systems, each in a different $^4$He cryostat, were employed for these low-temperature specific heat measurements, giving indistinguishable results. In the first system (system 1), a Lakeshore 336 Controller was used to measure both sample and reservoir temperatures. A Cernox 1030 sensor with temperature calibration down to 0.3 K and a 10 mg heater chip of $R = 1$ k$\Omega$ at room temperature are attached diametrically opposed on the disc. The thermometer in the reservoir was a standard Silicon diode DT 470 SD. A closed loop PID between the Silicon diode and a $R = 50$ $\Omega$ heater in the reservoir enabled the temperature control in the reservoir for the thermal relaxation method. A programmable current source Keithley 224 was used to apply the power to the heater, and a Keithley 2000 multimeter to monitor the voltage drop along the resistor in a 3-terminal scheme. A second experimental system (system 2) was also employed for some low-temperature measurements in the CeSb$_2$ crystal, where a Lakeshore DRC-91CA Controller was used to fix the reservoir temperature. The experimental cell was composed of a sapphire disc identical to the one described above, a CCS thermometer (carbon glass) and a $R = 1$ k$\Omega$ chip as heater. The thermometer was excited using a Keithley 224 current source, and the voltage drop was read with a Keithley 2000 multimeter. Heater excitation and read out was analogous to the one described for the first system.

The samples used in this work were cleaned using a scalpel to remove any possible oxide on the surface and then immediately fixed to the calorimeter using a small amount of low temperature grease. Closing and pumping the chamber down to $10^{-2}$ mbar was done within half an hour in order to avoid further oxidation of the samples. The amount of the sample employed for the measurements was 436 mg of $CeSb_2$, 252 mg of HOPG graphite and 292 mg of natural graphite, respectively. Sample placement was carefully chosen in the center of the triangle described by the sensors and thermal link anchoring in order to optimize internal equilibrium. In Fig. 1 a sketch of the experimental cell and the $^4$He insert is shown.

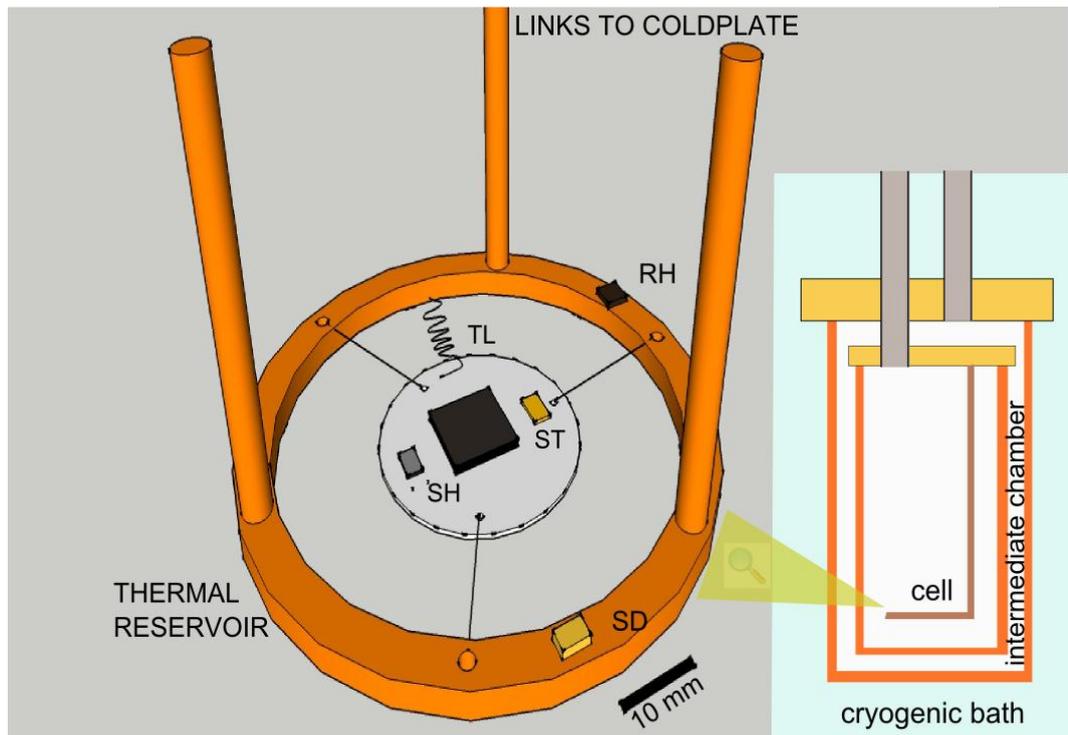

**Fig1** Calorimetric set-up for the low-temperature specific heat measurements. The sapphire disc is fixed using nylon wires in the middle of the copper ring (thermal reservoir). Sample heater (SH), thermometer (ST) and thermal link (TL) are placed forming a triangle. The sample is located in the center to optimize thermal equilibrium with the three elements described above. The reservoir temperature is controlled using a closed PID loop between a silicon diode (SD) and a heater (RH). Inset: Double-pot cryostat for calorimetric measurements at low temperatures

## 2.3 Methods

In this work two different calorimetric methods have been employed: (i) the thermal relaxation method, both the standard one and an alternative version developed in our laboratory [8] for longer relaxation times between cell and thermal sink (when typically above 3 minutes); and (ii) a *quasi-adiabatic* continuous method, previously implemented [8] and employed [21] for calorimetric characterization at and above liquid nitrogen temperatures.

In the standard relaxation method, the cell is driven from an initial equilibrium with the thermal reservoir to a stationary state at a higher temperature (typically $\Delta T_\infty /T \approx 1\%$) by an applied power $P$. Heating and cooling curves obey an exponential dependence with time, and are related to the heat conductance $\kappa_H$ of the thermal link (TL) and the relaxation time constant $\tau$, from which the total heat capacity $C_p(T)$ is easily derived as given by Eq (1) and Eq (2):

$$\kappa_H = P/\Delta T_\infty \quad (1)$$

$$C_P = \kappa_H \cdot \tau \quad (2)$$

where $\tau$ is obtained by fitting the relaxation exponential curve.

In the alternative relaxation method, the stationary state is not reached, but the power applied is switched off at some time during the heating process (usually when $\Delta T/T \approx 1\%$). Again the total heat capacity is calculated after obtaining $\tau$ from the relaxation curve, and $\kappa_H$ by a fit of the heating curves given by $\Delta T(t) = (P / \kappa_H) \cdot (1 - e^{-t/\tau})$. Using these two methods, specific-heat measurements in the temperature range from, say, 0.1 K up to 30 K can be easily performed, but the time elaspsed to complete the whole measurement in one sample can be above one week. An example of these two methods applied to the low-temperature specific heat measurement (using the above mentioned experimental system 1) in HOPG is shown in Fig 2.

In principle, the standard relaxation method (Fig 2a) is employed with lower heat capacity values, what implies smaller relaxation time constants (usually below 2 minutes), whereas the alternative relaxation method (Fig 2b) is applied when heat capacity increases more rapidly than the thermal link conductance (usually at higher temperatures) and consequently $\tau$ grows above several minutes. In both methods, the relaxation time constant $\tau$ is straightforwardly obtained from linear fits in semilogarithmic plots, such as those in Fig 2c for the corresponding points in Fig 2a and Fig 2b. The inset in Fig 2 shows the additional $T$ vs $(1 - e^{-t/\tau})$ plot to determine the value of $\Delta T_\infty$ in the alternative relaxation method. Notice that the excellent linearity in all these curves of Fig 2c is a clear evidence of the quality of the conducted experimental method and of the assumptions made.

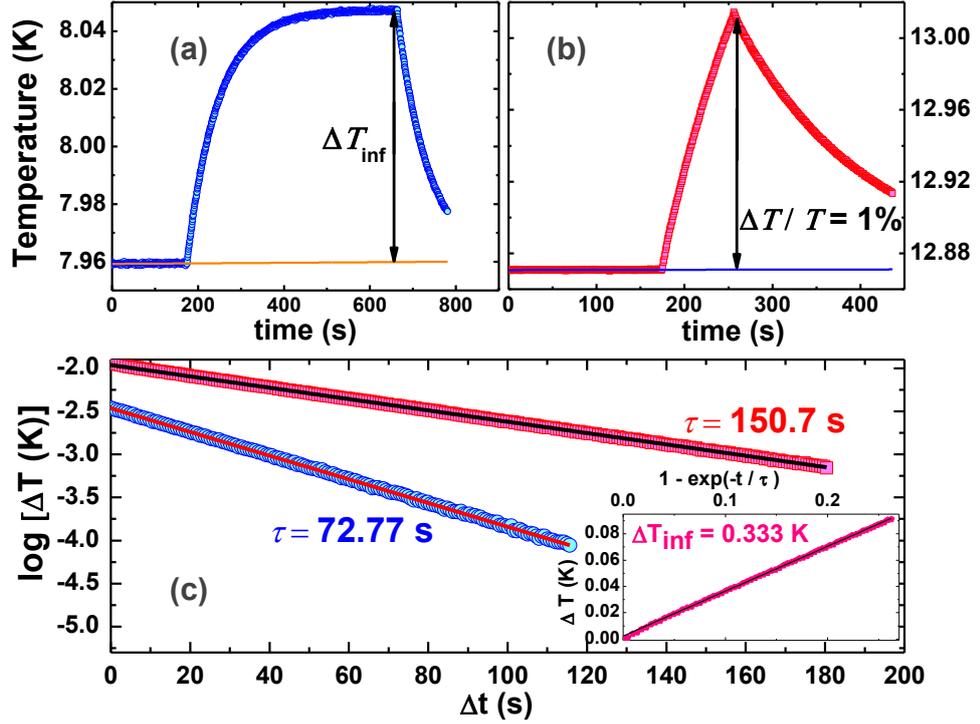

**Fig2** A real example of acquisition points (temperature versus time) obtained in HOPG when using (a) standard relaxation method and (b) alternative relaxation method. (c) Semilogarithmic plot for the corresponding points in (a) and (b) to extract the time constant $\tau$. Inset: Temperature increase versus $(1 - e^{-t/\tau})$ to determine $\Delta T_\infty$ in the alternative relaxation method.

A much faster calorimetric method, though usually less accurate, is the already mentioned continuous heating method [5]. The simplest and most traditional version is the ideal adiabatic one, where the simple equation $P = C_p \cdot (dT/dt)$ is assumed. This method has been usually employed at relatively high temperatures, though experiments down to 15 K have also been conducted [22]. An opposite version of this technique is exhaustively discussed in Ref. [9], where an accurate determination of the heat flux to the sample using commercial Peltier elements is performed. This version needs only a constant-reference temperature from the thermal bath, as well as a weak thermal contact between cell and bath ($\tau$ of the order of few tens of seconds). The use of a thermometer and a heater for the sample in four-terminal configuration is needed, in order to monitor the applied-power evolution with time/temperature. Heat losses calibration is performed in every experiment by simply recording the sample temperature evolution $T(t)$ with time by switching off the heating power at a given temperature $T(t_0)$.

We will discuss now our adapted version of the quasi-adiabatic continuous method [8], which we have used here for the first time at liquid-helium temperatures. As already described above, our calorimetric cell has a fixed weak thermal contact with the thermal sink, given by a copper wire (plus any other parallel thermal conduction mechanism that might contribute in a much lower amount) that results in relaxation time constants of the order of $\tau \sim 30$ s at liquid helium temperatures. The basic idea of the method is that there will always be an effective and measurable cooling power for the cell $P_{cool}(T)$ as a function of temperature for a given sample, *provided we keep fixed the temperature of the thermal sink*. All this means that the heat transport equation has to take into account both cooling and heating power terms:

$$C_\text{p}(T) \cdot \frac{dT}{dt}(T) = P_{heat}(T) + P_{cool}(T) = V_h(T) \cdot I_\text{h} + C_\text{p}(T) \cdot \Theta(T) \tag{3}$$

where $\Theta(T) = dT/dt$ accounts for the intrinsic *negative* thermal drift of the system measured by spontaneous or "standard" cooling ($I_\text{h} = 0$), this is, at zero applied heating power. The thermal sink is easily fixed (and monitored) at either 4.2 K (if helium bath is used), or 77 *K* (if liquid nitrogen, as in previous works [8,21]). The thermal drift of the system $\Theta(T)$ is determined for every single experiment, as it can vary depending on the total heat capacity of the cell (sample plus addenda). The heat capacity is hence determined by

$$C_\text{p}(T) = \frac{V_h(T) \cdot I_\text{h}}{\frac{dT}{dt}(T) - \Theta(T)} \tag{4}$$

Eq. (4) shows that, in order to calculate the heat capacity curve, both the heating *dT/dt* and cooling $\Theta(T)$ curves need to be combined together. Here comes a significant difference with other methods in the literature [9,22], where heating and cooling curves provide two *"independent"* heat capacity curves. Applied current to the heater $I_\text{h}$ is selected so that *dT/dt* is slow enough (typically < 3 K/min) to ensure good internal equilibrium in the cell. The empty-cell heat capacity is measured in a different run to subtract the addenda contribution.

At this point it is important to highlight the main difference that makes this method a fast and reliable tool to measure specific heat: whereas in the relaxation method the experimental time scale is governed by the thermal link between reservoir and cell, in the continuous method only internal thermal equilibrium in the cell is required to be fast compared to all other characteristic times, which at low temperatures is of the order of some tens of milliseconds. A thermal link between the cell and the sink is used so that the relaxation time constant $\tau$ is always of the order from half to one minute at liquid helium temperatures. This is a second important variance with earlier applied continuous methods, where thermal connection of the sample to the surroundings is chosen to be either extremely bad ($\tau \gg 1$ min) [5,22] or extremely good ($\tau \ll 1$ s) [9].

As can be seen, our low-temperature version of the quasi-adiabatic continuous method has something in common with the thermal relaxation one. An effective weak thermal link is needed to have an effective and measurable cooling rate given by the standard cooling $\Theta(T)$ (i.e., zero heating power), what allows us to employ *the same experimental set-up* for both thermal relaxation and quasi-adiabatic continuous techniques. Heating rates must be chosen carefully, usually below 2–3 K/min, to ensure that internal equilibrium is fulfilled at every single temperature among all the components in the experimental cell (see Fig 1). No limitation in heating or cooling as slow as possible exists, in principle. A remarkable aspect is that the density of data points in the heat capacity versus temperature plot is only conditioned by the maximal reading frequency of our electronics. This makes the method even more suitable for specific heat studies on systems with first order transitions, as the resolution of the peaks can be much better than with other methods. With the experimental cell described in this work, *dT/dt* on heating and cooling can also be directly used as valuable thermograms to monitor transitions such as those present in $CeSb_2$, as shown in Fig 3 (using the system 2).

To illustrate how our continuous method works, we shown in Fig. 3 different thermograms obtained to measure the specific heat of the $CeSb_2$ sample. The three magnetic transitions in $CeSb_2$ are clearly observed in the figure (on grey shadows), both in heating (Fig 3a) and cooling (Fig 3c) curves, hence also giving an estimation of the uncertainty in the determination of the transition temperature. On the other hand, *dT/dt* curves on heating for different applied currents (powers) (see Fig 3(a) and Fig 3(b)) can be used to check the reproducibility of the measurement. For the empty cell, heating (Fig 3(b)) and cooling (Fig 3(d)) smooth curves with no sharp features are observed, as expected. In fact, the heating and cooling rates used for the empty cell measurement are typically higher than those employed with the sample, due to its lower total heat capacity. These faster rates carry no problem, as internal equilibrium is more easily fulfilled for the addenda measurements.

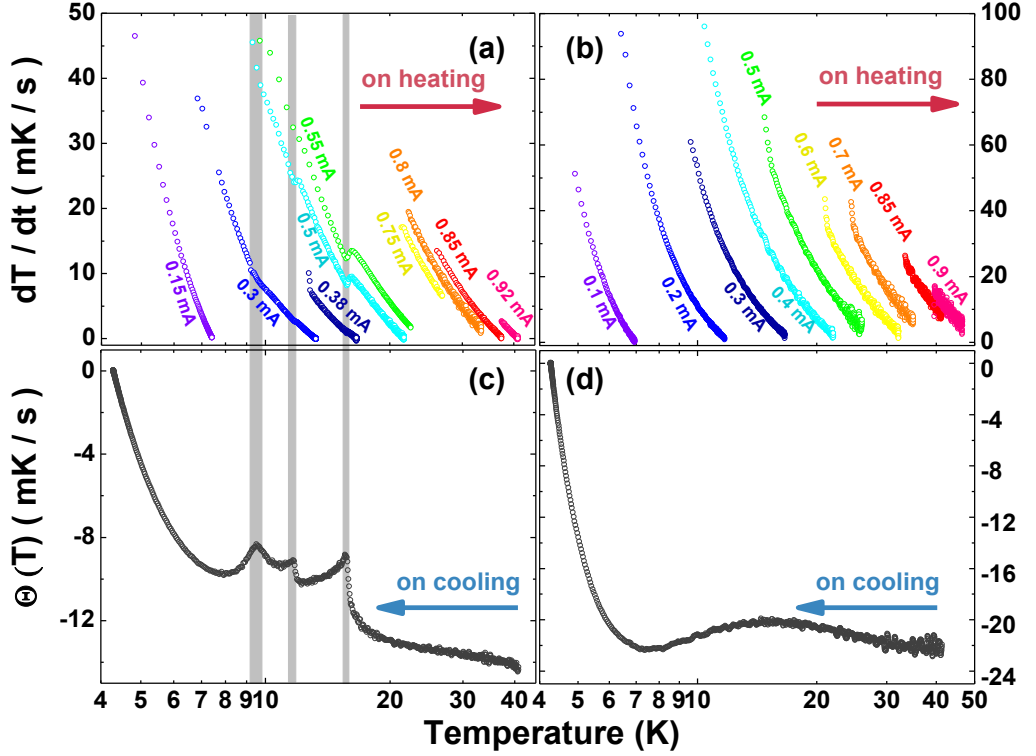

**Fig3** Temperature derivatives with time on heating $dT/dt$ [panels (a) and (b)] and on cooling $\Theta(T)$ [panels (c) and (d)]: (a) CeSb$_2$ and (b) empty cell, in both cases by heating with different applied currents; (c) CeSb$_2$ and (d) empty cell, by standard cooling at $I_h = 0$. The three magnetic transitions in CeSb$_2$ measurements [panels (a) and (c)] are highlighted with grey shadows at 9.5 K, 11.7 K and 15.6 K.

As described in detail in section 2.3, the heat capacity curve is determined via Eq. (4) from both the heating $dT/dt$ and cooling $\Theta(T)$ curves, taken under the very same experimental conditions, i.e., with the thermal sink at its fixed temperature. During the heating curve, the heater voltage is recorded as a function of temperature to take into account possible variations of its resistance with temperature. Although we usually measure the spontaneous cooling of the system $\Theta(T) = dT/dt$ at zero applied heating power ($I_h = 0$), Eq. (4) is also valid if cooling slower with a small applied current, by changing in the numerator the heating power by the net power difference.

Finally, to obtain the specific heat of the given substance, the heat capacity of the addenda is measured using the same procedure (e.g., panels (b) and (d) in Fig. 3) and then it is subtracted from the total heat capacity measured. For this case of the CeSb$_2$ sample, the corresponding heat capacity curves are shown in Fig. 4.

Moreover, we can see in Fig. 4 that the different heating runs, obtained using different currents, previously shown in Fig. 3 (a) and (b), merge very well into corresponding single curves of heat capacity, with differences well below 3% in absolute values.

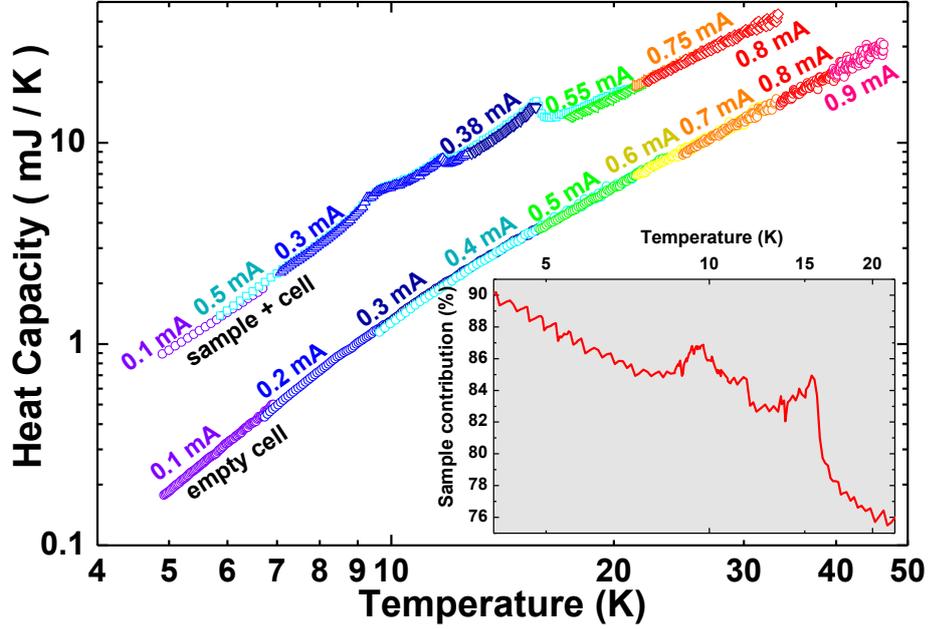

**Fig4** Heat capacity comparison among the different applied currents using the continuous method, calculated from the data in Fig 3 with Eq (4), for both the 'empty cell' and 'sample CeSb$_2$ + empty cell'. Inset: Relative CeSb$_2$ contribution to the total heat capacity.

From these curves, we can also assess the sample contribution to the heat capacity compared to the total one (sample + empty cell), as depicted in the inset of Fig 4. As can be seen, the sample contribution to the total heat capacity $C_{sample}/C_{total}$ varies from 75% up to the 88% in the region of interest for the magnetic transitions, what will make the specific heat absolute values well accurate (since the addenda contribution is small enough).

## 3 Results and Discussion

We have measured the specific heat of two different samples of great interest nowadays using a new low-temperature method, cross-checked against the results given by the well-established relaxation method and compared to previous work.

## 3.1 CeSb$_2$

In Fig. 5 we present the specific-heat data obtained for the CeSb$_2$ crystal, by using both the continuous and thermal relaxation methods. In the former case, eq. (4) was applied to the combined heating and cooling curves previously shown in the thermograms of Fig 2. After subtracting the cell contribution, we obtained the molar specific-heat data of this compound, shown in Fig 5.

As said above, we observe in Fig 5 that CeSb$_2$ exhibits three transition-like features, two of which have a $\lambda$ shape –those at 11.7 K and 15.5 K–, whereas at 9.5 K a round-shaped transition can be seen, in agreement with earlier experiments [11], reporting on those three ordered magnetic transitions at low temperatures and zero field. Measurements down to T = 2 K were performed and no further anomaly was found in the specific heat, in contrast to what was reported around 2.6 K in earlier experiments [11].

We want to emphasize that the very good agreement between the two calorimetric methods employed in the specific-heat measurement of CeSb$_2$ (each conducted in a different experimental set-up), including a good definition of the magnetic peaks in the specific heat, is a further evidence of the reliability of our quasi-adiabatic continuous method at low temperatures.

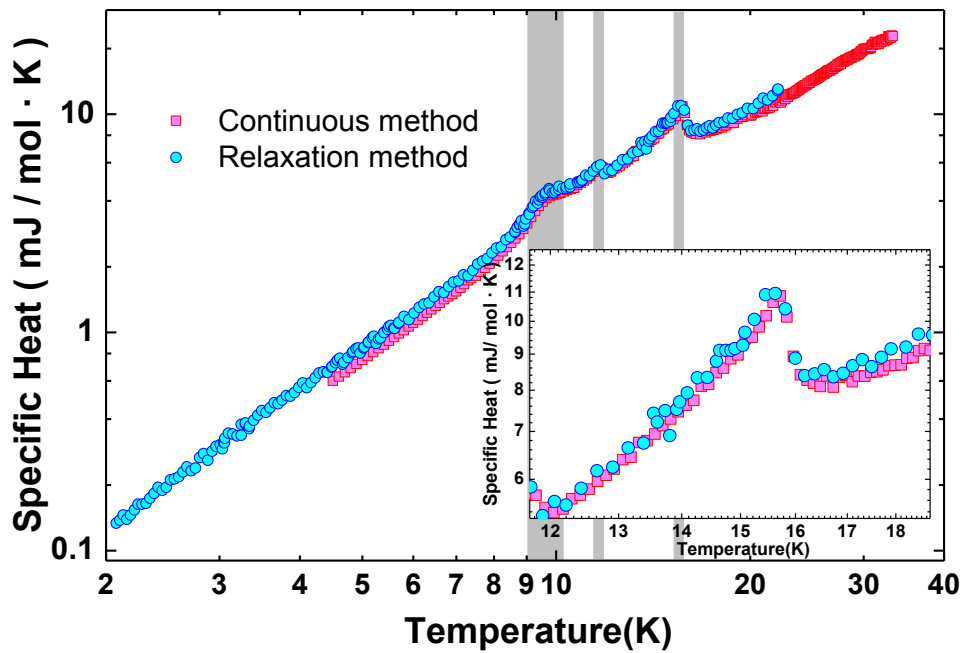

**Fig5** Specific-heat values for CeSb$_2$ with the three magnetic transitions at 15.5 K, 11.7 K and 9.5 K marked on grey shadows. Blue circles correspond to the measurement using the relaxation method (both standard and fast one), red squares account for the fast quasi-adiabatic continuous method. Inset: zoom on the 15.5 K transition, where the difference between both methods is found to be < 3%. Experimental system 1 was used for the relaxation measurements, whereas the quasi-adiabatic continuous method was employed in system 2.

### 3.2 Graphite

Low-temperature specific heat on two different graphite samples has been measured down to $T = 2$ K to look for possible different behavior in natural graphite (with an expected higher degree of crystallinity) compared to HOPG graphite, due to their different density of interfaces or defects given by their different monocrystals size. We have also performed a low-temperature study for HOPG using the relaxation as well as the quasi-adiabatic continuous method in order to check again any possible differences, in case systematic errors may appear. As can be seen in Fig. 6, no significant difference is found between the relaxation and the quasi-adiabatic continuous methods, as differences in absolute value keep always below 3%, which is essentially the experimental error. In addition, earlier published data of another HOPG sample (from Union Carbide Co.) by Alexander et al. [23] in an even wider low-temperature range are also shown in Fig 6. As can be observed, a very good agreement is also found between our data and those from [23]. Therefore, we can conclude that using our quasi-adiabatic continuous method at low temperatures is well justified.

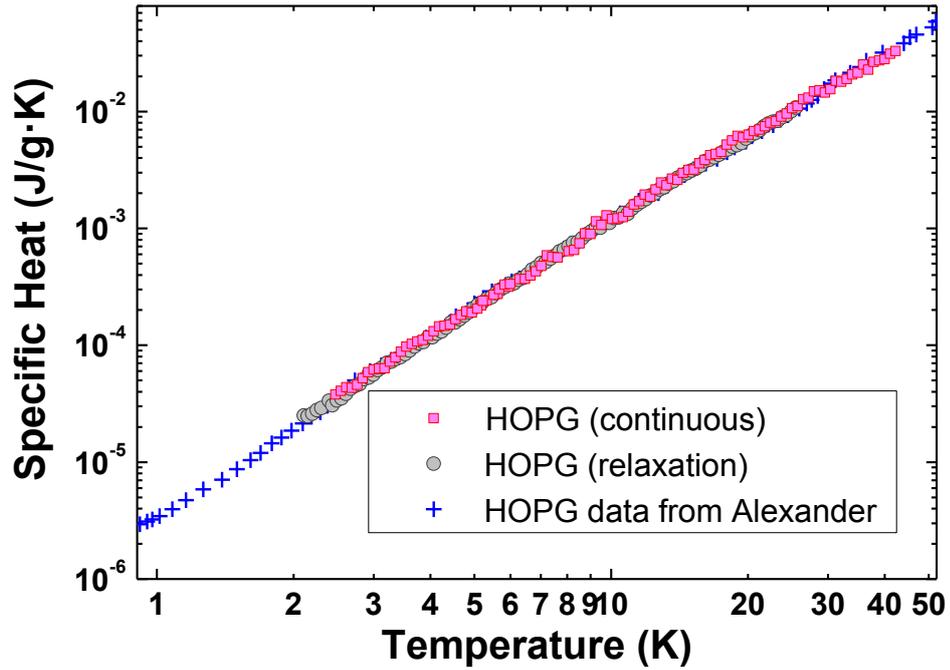

**Fig6** Comparison of specific-heat data obtained for the HOPG sample using either thermal relaxation (circles) or quasi-adiabatic continuous (squares) methods, employing in both cases the experimental system 1. Earlier published data of another HOPG sample by Alexander et al. [23] are also shown (crosses).

In Fig 7 we compare our obtained specific-heat data at lower temperatures, both for HOPG and natural graphite, to abovementioned data of another HOPG sample [23], as well as to earlier published data below 1.9 K of a natural Madagascar graphite (NMG) by van der Hoeven, Jr. and Keesom [17], and of spectroscopic pure graphite powder by Mizutani *et al.* [18]. The inset shows the same $C_p$ data from Refs. [17] and [23] only in the range 0.4 K–1.4 K, as well as their corresponding least-squares fit to obtain the expected linear (electronic) and cubic (Debye) contributions to the specific heat. One finds that the usual $C_p = \gamma T + c_D T^3$ behavior is really valid only below 1.2 K, where van der Hoeven, Jr. and Keesom [17] obtained $\gamma = 13.8 \mu J/mol \cdot K^2 = 1.15 \mu J/g \cdot K^2$, and $c_D = 27.7 \mu J/mol \cdot K^4 = 2.31 \mu J/g \cdot K^4$, *i.e.* a Debye temperature $\Theta_D = 413$ K, and Alexander et al. [23] obtained $\gamma = 17.3 \mu J/mol \cdot K^2 = 1.44 \mu J/g \cdot K^2$, and $c_D = 24.9 \mu J/mol \cdot K^4 = 2.07 \mu J/g \cdot K^4$, *i.e.* a Debye temperature $\Theta_D = 427$ K . In contrast, Mizutani *et al.* [18] found that their measurements in pure graphite between 1.5 and 4.2 K contrarily followed well $C_p = \gamma T + c_D T^3$, with $\gamma = 30 \mu J/mol \cdot K^2$ and $c_D = 26 \mu J/mol \cdot K^4$, *i.e.* a similar Debye temperature $\Theta_D = 421$ K.

Our measurements show specific-heat data for our natural graphite sample slightly higher than those of HOPG. In any case, both samples clearly show a behavior deviating from a linear $C_p = \gamma T + c_D T^3$ curve by below, very much in agreement with the another HOPG [23] and the NMG data [17], but significantly lower at $T > 2.5$ K than those reported in pure graphite by Mizutani *et al.* [18].

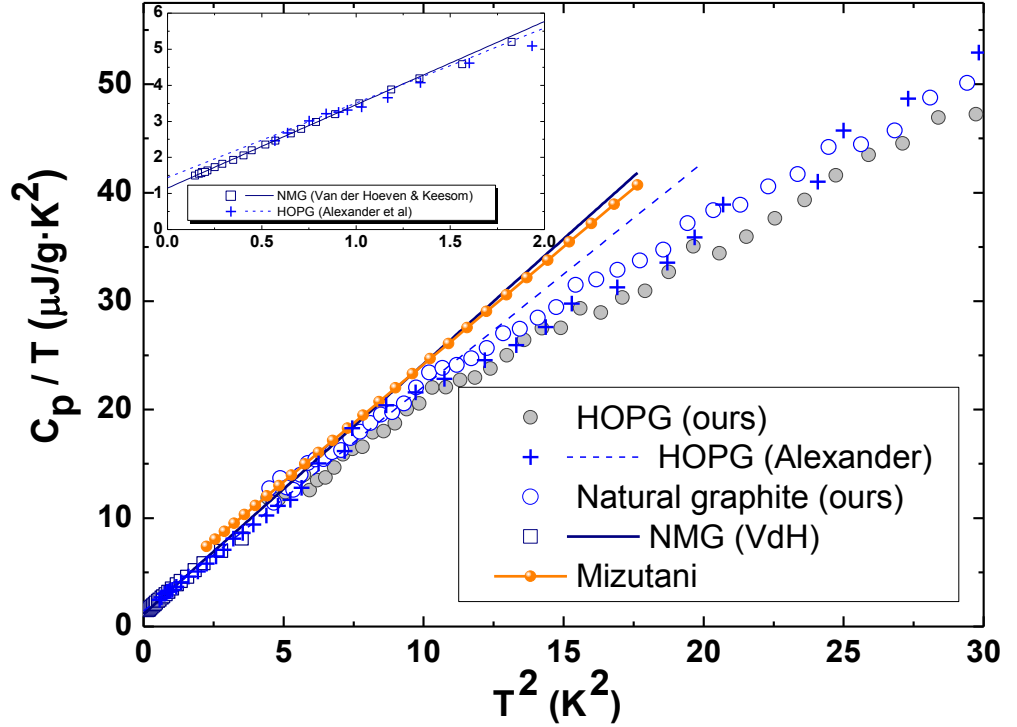

**Fig7** $C_p/T : T^2$ plot of our low-temperature specific heat data obtained for both HOPG (solid circles) and natural graphite (open circles), compared to earlier data of another HOPG sample by Alexander et al. [23] (crosses), of a Madagascar natural graphite (open squares) by van der Hoeven, Jr. and Keesom [17], and of spectroscopic pure graphite powder by Mizutani *et al.* [18] (linked balls). Inset: enlargement of the lowest temperature data in the range 0.4 K – 1.4 K from van der Hoeven, Jr. and Keesom [17] and from Alexander et al. [23], with their corresponding least-squares fits (thin and dashed lines, respectively) to obtain the linear and cubic coefficients of $C_p = \gamma T + c_D T^3$, found to be valid only below 1.2 K.

In Fig 8 we present these and other specific-heat data in a wide log-log scale. It is interesting to compare again our HOPG and natural graphite data with earlier data on HOPG by Alexander et al. [23], but also with different qualities of graphite samples also measured by van der Hoeven, Jr. and Keesom [17] (see the different symbols indicated in the legend). As can be more clearly seen in Fig 1 of Ref. [17], there is a systematic increase in the low-temperature specific heat of graphite with increasing disorder, respectively: NMG (very low degree of stacking faults), pile graphite (low degree of stacking faults), graphitized lampblack (high degree of stacking faults), and pyrographite (very high degree of stacking faults). As could be expected, both HOPG and natural graphite specific-heat curves tend to values corresponding to a low degree of stacking faults.

We also include in Fig 8 theoretical calculations [24] of the specific heat from the vibrational density of states for graphene, a single-walled nanotube (SWNT), a nanotube rope and pure graphite (respectively shown by different lines from top to bottom, as indicated in the legend). Possible linear ($\propto T$) or quadratic ($\propto T^2$) dependences of the specific heat are also indicated by labelled solid lines.

Much discussion can be found in the literature [17,23,24] about the expected temperature dependence of the specific heat of pure carbon as a function of the *dimensionality* of the material, when going from 3D graphite to 2D graphene and to 1D single-walled nanotubes (SWNT).

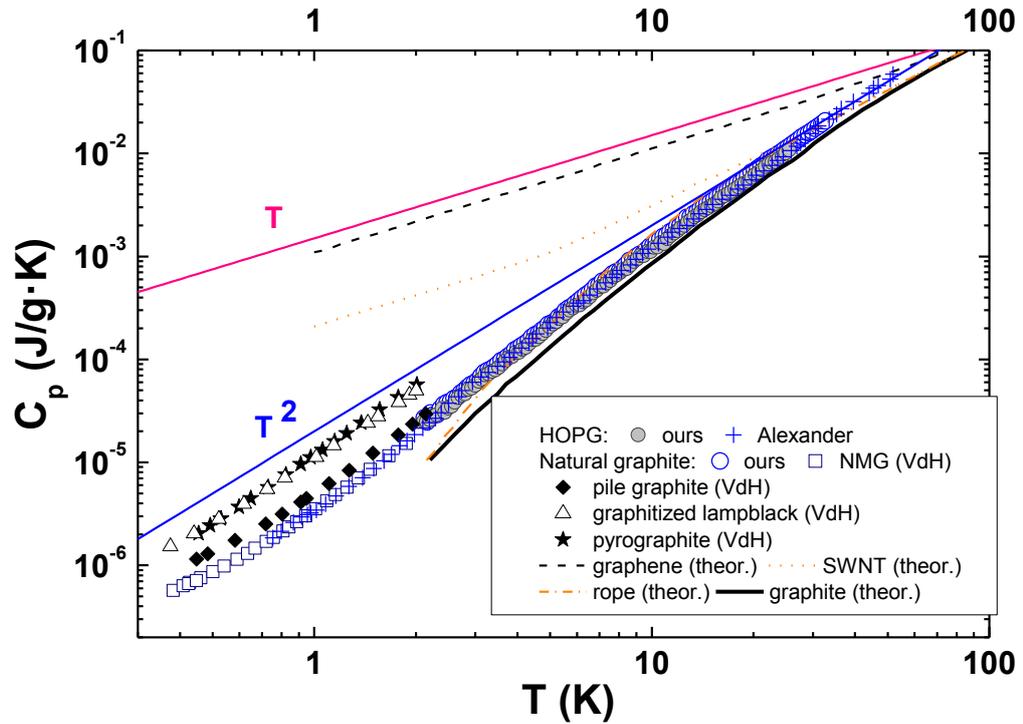

**Fig8** Low-temperature specific heat data in a log-log scale from different sources: comparison of our HOPG data (filled grey circles) and natural graphite data (open circles) with an earlier measurement on HOPG by Alexander et al. [23] (crosses), with different qualities of graphite samples by van der Hoeven, Jr. (VdH) and Keesom [17] (different symbols as indicated in the legend), and with theoretical calculations [24] for graphene, a single-walled nanotube (SWNT), a nanotube rope and pure graphite (different lines as indicated in the legend). Possible linear ($\propto T$) or quadratic ($\propto T^2$) dependences of the specific heat are suggested by labelled solid lines.

In principle, a non-negligible electronic contribution arising from the density of states at the Fermi level is only expected in 3D graphite. A linear electronic term of $\gamma = 12.6 \mu J/mol \cdot K^2$ was calculated [17], in very reasonable agreement with the above reported experimental values for pure graphite.

The main problem is to determine and understand the contribution of acoustic phonons to the specific heat. First of all, let us remember that an acoustic phonon branch in $d$ dimensions with a dispersion curve $E(q) \propto q^\alpha$ will give $C_p \propto T^{d/\alpha}$.

As thoroughly discussed by Hone *et al.* [24], in an isolated SWNT all of the circumferential degrees of freedom are frozen out at low temperature and the phonons are therefore strictly 1D, with four acoustic branches. Hence $C_p$ is predicted to be linear in $T$ at the lowest temperatures, with an increase in slope due to the contribution of the first acoustic subband above $\approx$ 5K, as can be seen in Fig 8 (dotted line). In the opposite case of a bundle of strongly coupled nanotubes (a nanotube "rope"), the calculated curve (dash-dotted line in Fig 8) diverges below the SWNT $C_p(T)$ curve for $T < 30$ K, following a curve similar to that of 3D graphite. Experimental $C_p(T)$ curves of SWNT bundles [24,25] show good agreement above $\approx$ 4–5 K with the calculated ones for isolated nanotubes, but the expected linear regime at low temperature is never reached. For instance, Bagatskii *et al.* [25] obtained a power law $C_p(T) \propto T^{1.73}$ below 5 K.

In contrast to the 1D phonon density of states of the nanotube, that of a single 2D sheet of graphene is much greater in magnitude at low energies, because a graphene sheet is weak to bending. This acoustic "layer bending" branch in graphene [24] behaves quadratically as $E(q) \propto q^2$, instead of the typical linear dispersion curve found in 3D solids. As a result, the calculated $C_p(T)$ curve of graphene has a roughly linear temperature dependence $C_p(T) \propto T$ in a wide temperature range, and is expected to be much larger than that of 1D SWNT and even much more than 3D graphite. Unfortunately, no experimental $C_p(T)$ data on graphene are available to check this.

Finally, the $C_p(T)$ curve of graphite calculated from the vibrational density of states is also shown in Fig 8 (lower solid line). The coupling between adjacent graphene sheets by passing from 2D graphene to 3D graphite, translates into a perpendicular phonon dispersion which shifts the phonon spectral weight to higher energies [24], hence producing a strong decrease in the specific-heat magnitude, as well as eventually a cubic temperature dependence at low temperatures. In fact, the crossover from the 3D behavior of graphite at low temperatures to a more 2D behavior at higher ones, is calculated to occur at around 100 K [24]. Experimentally (see Fig 8), we find that even the purest graphite samples exhibit $C_p(T)$ curves well above the calculated one and with temperature dependences weaker than a Debye-like $C_p(T) \propto T^3$, but still always stronger than a mere $C_p(T) \propto T^2$ dependence. Even removing the electronic linear contribution, neither a cubic nor a quadratic neat contribution can be found in any temperature range.

## 4. Conclusion

A new, fast quasi-adiabatic continuous method has been presented for low-temperature specific heat studies. Comparison to a widespread used method such as the thermal relaxation one, has allowed us to test the validity of our new method, which has demonstrated a good accuracy even in specific-heat absolute values. The interplay between these two methods requires no extra effort in the experimental set-up described here, as both of them can be used by only changing the data acquisition software. The strong point of the new implemented method lays on the speed to measure the specific heat of samples in an accurate and exhaustive way, making it possible to reduce the experimental time from several days to a few hours. This is accompanied by a high-density of data as experimental output, ideal to resolve phase transitions in superconductors or ferro/antiferromagnetic compounds. This will allow us to undertake studies in different families of compounds with much lower time consumption, and, as a consequence, far lower liquid-helium requirement.

These features presented above come up to make this method a good alternative to commercial standard systems such as PPMS (Physical Properties Measurement System), where relaxation techniques are used for specific-heat measurements.

At the same time, we have presented in this work an exhaustive study on the specific heat between 2 and 40 K of two different materials of interest, namely $CeSb_2$, from the family of the light rare-earth diantimonides, and two different graphite specimens (natural and HOPG).

In the case of the $CeSb_2$ crystal, our low-temperature specific heat measurements have shown three maxima associated with magnetic transitions at zero field, two of which with a $\lambda$ shape –at 11.7 K and 15.5 K–, and another round-shaped one at 9.5 K, all of them in agreement with earlier experimental reports [11].

We have also measured the specific heat of HOPG and natural graphite. We have found good agreement in both cases when comparing to other data found in the literature of *pure* graphite, whereas more disordered graphite or 2D nanotubes exhibit higher specific-heat curves. In contrast to some statements found in the literature, we observe neither a cubic nor a quadratic dominant contribution for the specific heat of graphite in any temperature range, with the only apparent exception below 1.2 K due to the dominance of the linear electronic contribution.

## 5. Acknowledgements

The Laboratorio de Bajas Temperaturas (LBT-UAM) is an associated unit with the ICMM-CSIC. This work was partially supported by the Spanish MINECO (FIS2011-23488, and Consolider Ingenio Molecular Nanoscience CSD2007-00010 program) and by the Comunidad de Madrid through program Nanobiomagnet (S2009/MAT-1726). T. P.-C. acknowledges financial support from the Spanish Ministry of Education through FPU grant AP2008-00030 for a PhD thesis. J. H. acknowledges financial support from the Spanish Ministry of Education through grant SB2010-0113 for a postdoctoral stay. We are grateful to Daniel Farías for providing us with the sample of natural graphite, and to Paul Canfield for his stay at our Laboratory where he led the implementation of an experimental set-up for growing metallic crystals at high temperatures, within our MSc program on Condensed Matter Physics and Nanotechnology. The $CeSb_2$ crystal was grown by us using that system.